\parindent=0pt
\magnification=\magstep1
\line{MNRAS, in press\hfill}
\bigskip
{\bf A robust method for investigating galactic evolution in the
submillimetre waveband: II the submillimetre background and source
counts}

\bigskip
{\bf Stephen A. Eales\footnote\dag{\rm e-mail: sae@astro.cf.ac.uk} and 
M. G. Edmunds\footnote\ddag{\rm e-mail: mge@astro.cf.ac.uk}}

\bigskip
Department of Physics and Astronomy, University of Wales, Cardiff, P.O. Box
913, Cardiff CF2 3YB.

\bigskip

\centerline{\bf Abstract}

\parindent = 20pt
\bigskip
This is the second of two papers describing a model of galactic evolution in
the submillimetre waveband. The model incorporates a self-consistent treatment
of the evolution of dust and stars, is normalized to the submillimetre
properties of galaxies in the local universe, and can be used to make predictions for 
both disk and elliptical galaxies and for
`closed-box', `inflow', and `outflow' models of galactic evolution. 
The model does not include the effects of hierarchical clustering,
but we show that the variation in the predictions produced by
the different dust-evolution models is
so large that it is premature to include 
the effects of an even more uncertain process.
In Paper I we investigated whether it is possible to explain the extreme
dust masses of high-redshift quasars and radio galaxies by galactic
evolution.
In this paper we use the model to make
predictions of the submillimetre background and source counts. 

All
our disk-galaxy models exceed at short wavelengths ($\rm \lambda < 200
\mu m$) the
submillimetre
background recently measured by Puget et al. (1996), suggesting that
either there is a problem with our models, or that the background
measurement at the shorter wavelengths is inaccurate, or a combination 
of the two. However, the two models in which we assume the 
rapidly-evolving star-formation rate found from optical studies
predict backgrounds that are so much greater than the measured background
that we do not believe that this discrepancy can be due to
an inaccurate background measurement. We therefore conclude that
there is no evidence in the submillimetre waveband for the
rapid evolution found from optical studies.
At longer wavelengths, most of the disk-galaxy models predict
backgrounds that are less than the observed background.
Our elliptical-galaxy models are more uncertain because they are
less securely tied to the observed submillimetre properties of the
local universe. We find that it is relatively easy to produce models
that predict backgrounds similar to that observed, but we caution
that our disk-galaxy models show that a significant fraction of the
observed background must be coming from disk galaxies.
The biggest weaknesses in our models are the
lack of a direct measurement
of the submillimetre luminosity function for galaxies
and the fact that we have been forced to assume that all local
disk-galaxies have the same far-infrared---submillimetre spectral
energy distribution and the same ratio of gas mass to stellar mass.
Observations with the SCUBA submillimetre array should remove both
of these weaknesses, as well as providing measured source counts with
which to confront the models.

\bigskip

\vfill
\eject
\centerline{\bf Introduction}

\parindent = 20pt

\bigskip

The submillimetre waveband ($\rm 100\ \mu m < \lambda < 1\ mm$) is one
of the few electromagetic wavebands that has not been used
extensively for astronomical observations. 
In particular, our knowledge of the submillimetre properties
of galaxies is very poor, with there only being a
handful of submillimetre flux measurements of galaxies
(Eales, Wynn-Williams \& Duncan 1989; Stark et al.
1989). Our ignorance is
due partly to
there only being a few atmospheric windows within this waveband and
partly to technological limitations: continuum observations
in this waveband have relied on single bolometers,
which means that only a single point in the sky can be observed at
one time. However, the situation is rapidly changing. As we write,
the Infrared Space Observatory (ISO) is making observations at
wavelengths up to 200 $\mu$m, thus covering the short-wavelength end
of the submillimetre
waveband, and an array of bolometers (the {\bf S}ubmillimetre {\bf C}ommon
{\bf U}ser {\bf B}olometer {\bf A}rray---SCUBA [Duncan 1990]) is being 
installed
on the James Clerk Maxwell Telescope. 
The advantage of the latter instrument is that, first, of course, arrays
make
it possible to map the sky much more quickly but, second, and perhaps
more important, the individual bolometers in SCUBA are ten times more sensitive
than previous bolometers. Finally, there is an instrument that is
no longer operating: the Cosmic Background Explorer Satellite.
COBE was designed to measure the integrated infrared and submillimetre
emission from galaxies, and recently Puget et al. (1996) have claimed
to find, in the vast COBE database, evidence for this emission.

The submillimetre waveband is important to our understanding of galaxies
for two reasons. The most basic reason is that galaxies do
emit a significant fraction of their emission in this waveband (Eales,
Wynn-Williams \& Duncan 1989; Stark et al. 1989), so a proper
energy budget for individual galaxies is impossible without submillimetre
observations. The second more important reason is that the presence
of dust means our view of the galaxy population as a whole
is
always biased at optical wavelengths---a 
bias which can be corrected by observing the
submillimetre emission from the dust that is doing the obscuring.
An example of this is the question of the origin of elliptical galaxies.
Since there are large numbers of old stars in these objects but very
little current star formation, it seems likely that the stars were
formed over a relatively short period. However, searches at optical
wavelengths have failed to find any evidence of a cosmic epoch in which
this star formation is occurring (e.g. De Propris et al. 1993).
A possible solution to this problem is if the galaxies in this phase
are shrouded with dust: in which case the galaxies will be submillimetre
sources (Bond, Carr \& Hogan 1986).

There has been a large number of attempts to predict the submillimetre
background and source counts that should be produced by the `universe
of galaxies'. Accurate predictions require knowledge of two things 
which are not well known: the submillimetre properties
of the local universe and how galaxies evolve. Previous models of the background
and the source counts generally fall into two groups.
In the first fall those which are tied closely to observations
but make no attempt to incorporate a physical model for how galaxies
evolve. For example, Beichman \& Helou (1991) and Eales (1991)
based their models on the local far-infrared luminosity function and on
observational evidence about the spectral energy distributions of galaxies
in the far-infrared--submillimetre waveband but made no attempt to
incorporate, our admittedly limited, understanding of how galaxies evolve;
instead they just assumed simple functional forms for the evolution and
investigated how the backgrounds and the counts changed as the parameters
of these functions were changed. Our
criticism of these models is that we do know {\it something} about the
physics of galaxy evolution and this should be incorporated in the
models.
In the other camp are the 
models which incorporate detailed physical models for how galaxies
evolve (Wang 1991a,b; Franceschini et al. 1994). Our minor
criticism of these models is that
the treatment of what is known about the submillimetre properties
of galaxies tends to be rather cursory.
Our major criticism is
that they tend to be very complex with large numbers of parameters, which
creates a number of related problems. 
The complexity means that it is usually unclear how critically the
predictions depend on the accuracy of the assumptions of the model.
Wang's models (Wang 1991a), for example, contain assumptions about how dust
is created and destroyed in the interstellar medium, processes which
are still poorly understood (Whittet 1991), and it is unclear how
critically the model predictions depend on these assumptions. 
From a practical observer's point of view, one would like to be able to
see which are the critical assumptions in any model and whether these
are susceptible to observational confirmation.

Encouraged by the new observational opportunities,
we have started a project to model galaxy 
evolution in the submillimetre waveband. Our project has three
goals. We want our models to incorporate all that is known about
galaxies in this waveband and, as the observations improve, we want
it to be easy to include the new information. We want to incorporate
a physical model for how galaxies evolve, but one which, while covering
as wide a variety of types of evolution as possible, is also fairly
simple. Lastly, we want to be able to determine the critical
assumptions on which the model is based and whether these can be
improved by future observational or theoretical work.
Given the new submillimetre instruments, which will make
all kinds of new observational projects possible, this final point
is an important
one. As an example, we find that, as expected, the strength of the
submillimetre background can be used to distinguish between
types of galactic evolution, but that our predictions depend critically
on our assumptions about the submillimetre properties of nearby galaxies:
and thus, something which is not obvious, a key cosmological project 
to carry with the new submillimetre
arrays is actually to observe large numbers of nearby galaxies. 

There are two groups who have carried out projects with
similar objectives to ours. Blain \& Longair (1993, 1996)
have investigated the effect of the evolution of large-scale structure
on the submillimetre background and source counts. We decided not to include
the effects of this in our models, partly because there is no consensus
about any aspect of the evolution of large-scale structure, but mainly
because we find the uncertainties in the evolution of dust in individual galaxies
(something not considered by Blain \& Longair) lead to a large spread in the predicted
background and counts, without including the much more uncertain process of large-scale
structure evolution.
Fall, Charlot \& Pei (1996) have recently presented models 
of the evolution of Ly$\alpha$ absorption systems which are an
elegant synthesis of our current ideas about galactic evolution
and of what is known about observationally about these objects. 
These models produce estimates of the backgrounds in all wavebands.
Our approach is quite different from that of Fall et al., since we
discard generality and elegance, and aim to produce a very simple model,
as closely linked to the present submillimetre observations as possible, and
one possible to easily modify when new submillimetre data arrives.

In the first paper (Eales \& Edmunds 1996; henceforth EE), 
we used
our model to show that galactic evolution can account for
the high dust masses of a handful of high-redshift
quasars and galaxies but only for a
very restricted range of model parameters.
In this paper we use our models
to make predictions for the submillimetre background and source counts. 
The details of the models are in \S 2. 
The results and discussion are
in \S 4.
Since we do not want to interrupt the flow of the paper
to analyse all the uncertainties in the models, we have
noted every place where we think there is some uncertainty by
a `U' in parentheses, with usually a short discussion in a footnote.
Finally, we note that as one of our goals is to give observers a relatively straightforward
theoretical framework with which to start to interpret the avalanche of submillimetre observations
of galaxies that is (hopefully) about to occur, copies of the software used in this paper,
which can be used to generate backgrounds and source counts at any wavelength
can be obtained from the authors.
\bigskip
\centerline{\bf 2. The Models}
\bigskip
\centerline{\bf 2.1 Models of how the dust mass evolves}
\bigskip

Our models are based on the fundamental assumption 
that the mass of dust in a galaxy is always proportional
to the mass of metals in the interstellar medium. The
observational evidence that this is true at both high and low redshift
is fairly good and is given in EE (U\footnote\dag{Our models do not
rely on us knowing the absolute value of the fraction of metals bound
up in dust, merely that this ratio does not change with redshift. If
this ratio is different by, for example, a factor of two at high redshift
than it is at low redshift, our predicted submillimetre fluxes for
the high-redshift objects will be off by the same factor. For further
discussion, see EE.}). The advantage of this assumption is that we can
sidestep the many uncertainties about the properties of dust.
A second less fundamental assumption, which we make
for convenience, is that there is no significant
delay between the formation of a coeval population of stars and the
production of dust by this population (see EE for a discussion).

Once we have made these assumptions, we can use standard chemical
evolution models, for which we use the notation of Edmunds (1990). We 
assume (like Fall, Charlot \& Pei [1996]) that the dust mass
$M_d$ present in a galaxy is simply proprtional to the product
$zg$ of the mass fraction $z$ of heavy elements in the interstellar
medium with the gas mass $g$. We allow inflow of unenriched gas into
the system or outflow of gas at the ambient metallicity, and use
the simplest `linear' models (e.g. Edmunds 1990, and references therein)
in which the inflow or outflow rate is linearly proportional to the
star-formation rate (SFR: $\rm inflow = \gamma \times SFR$ or
$\rm outflow = \lambda \times SFR$). Although real galaxies will have
more complicated variations of inflow or outflow, these models should
give a reasonable idea of the likely effects of different types of
chemical evolution, and have the advantage that the time variation
of the star formation need not be made explicit. The total
mass $M_{tot}$ of the galaxy at any time is the sum of the
gas mass $g$ (which includes the dust mass) and the stellar mass $\alpha s$,
$s$ being the total mass of stars that have formed and $\alpha$ being
the mass fraction of the stars 
which remains locked up in long-lived
low-mass stars or stellar remnants (we assume a value for $\alpha$ of
0.8). The
gas fraction $f$ is defined as gas mass/total mass. It can be shown that,
for the three different scenarios of galactic evolution---closed-box,
inflow, and outflow---and starting from an initial unit mass of gas, the
following equations hold:
\medskip
\parindent = 0pt

\line{\it Closed-box:\hfill}
\medskip
$$
M_d = k f ln(1/f),
$$
\medskip
\noindent in which $k$ is the product of the yield (Edmunds 1990)
and the fraction by mass of the metals in the interstellar medium
that are incorporated in dust.
\medskip

$$
\alpha s = 1 - f,
$$
\medskip

$$
M_{tot} = 1.
$$
\medskip

\line{\it Outflow:\hfill}
\medskip

$$
M_d  = {kg ln(1/g) \over 1 + \lambda / \alpha},
$$

\medskip

$$
g = {f \over 1 + (\lambda/\alpha) (1 - f)},
$$

\medskip

$$
\alpha s = {1 - f \over 1 + (\lambda/\alpha) (1 - f)},
$$
\medskip

$$
M_{tot} = {1 + (\lambda/\alpha)g \over 1 + (\lambda/\alpha)}.
$$

\medskip
\line{\it Inflow:\hfill}
\medskip

$$
M_d = {kg \over \gamma/\alpha} (1 - g^{(\gamma/\alpha)/(1-\gamma/\alpha)} ),
$$
\medskip

$$
g = {f \over 1 - \gamma/\alpha(1 - f)},
$$
\medskip

$$
\alpha s = { 1 - f \over 1 - (\gamma/\alpha) (1 - f) },
$$

\medskip

$$
M_{tot} = {1 - (\gamma/\alpha)g \over (1 - (\gamma/\alpha)}.
$$
\medskip

\parindent = 20pt

A way of getting insight into these equations is to normalize them
to the properties of low-redshift galaxies. Low-redshift spirals
typically have values of $f$ of about 0.1 (Young \& Scoville 1991). 
Figure 1 shows the ratio of the dust mass at a gas fraction $f$ to
the dust mass at a gas fraction $f=0.1$ i.e. the current value.
When a galaxy is formed it contains no stars and so $f=1$.
When star formation starts, metals, and thus
dust, start to be created, leading to a gradual increase in the total
dust mass. However, star formation also consumes gas and the dust which
is assumed to be intermingled with the gas. Eventually this second
process becomes the most important one, leading to a decrease in the
dust mass. Including inflow or outflow in the model changes the relative
efficiencies of the two processes. 
The rather surprising conclusion
from these curves is that the dust masses of spiral galaxies are
unlikely to have been significantly higher in the past than they
are today. Only
the outflow models produce dust masses
in the past substantially higher than those
today, and the largest increase in dust mass in Fig. 1 is probably
unrealistic because, as we argue in EE, a value for $\lambda/\alpha$
of 10.0 leads to too low metal abundances for present-day galaxies.
Thus the largest increase in the dust mass that is possible, even
when outflow models are considered, is $\rm \simeq 4$.

Since cosmic time does not come into these equations, these models
can be applied to ellipticals as well as to spirals, although as
the current value for $f$ of ellipticals is much less than 0.1,
the curves in Fig. 1 are not particularly useful. Once a star-formation
history is assumed for a galaxy, the equations can be used to calculate
how dust mass depends on cosmic time. If the bulk of the star formation
in an elliptical galaxy does occur over a short period (\S 1), the
evolution of the dust mass seen in Fig. 1 will occur but over this short
period, with the maximum dust mass being the same as the maximum
dust mass of a spiral of the same total mass.
\bigskip

\centerline{\bf 2.2 Extending the models to submillimetre luminosity}

\bigskip

Once a history of star formation is assumed for a galaxy the change
of the dust mass with time can be calculated using the equations
above. However, this is not enough to predict how the
submillimetre luminosity depends on time. To predict this, we have
to make answer two questions: (1) What are the stars that are heating
the dust? (2) Is the dust in a galaxy absorbing most of the optical and
ultraviolet light from the stars (the dust is optically-thick) or is
only a small fraction of this light being absorbed (the dust 
is optically-thin)?

There is
a long-running
argument in the literature as to whether the bulk of 
the far-infrared emission from
galaxies
is from dust heated by old low-mass stars or from high-mass stars (e.g. Helou
1986, Persson \& Helou 1987; Boulanger \& Perault 1988; Devereux \& Young 1990)
which
seems to have petered out through exhaustion rather than consensus. 
We shall assume that the far-infrared and submillimetre emission is from
dust that is heated by young massive stars. The reason for this assumption
is the practical one that we can then link the far-infrared-submillimetre
properties of a galaxy directly to the current star-formation rate in the
galaxy. This assumption may not be correct in present-day 
spirals with relatively low star-formation rates, but for the majority
of the models in which we assume that the star-formation rate is constant
with time this actually doesn't matter (see below); and for the 
models in which the star-formation
rate was higher in the past, even if the assumption is not completely correct
at low redshift, it will be increasingly good as one moves to a higher
redshift.

For a typical starburst galaxy, a much larger fraction of the 
bolometric luminosity
is coming out in the far-infrared---submillimetre wavebands than in the
optical waveband (Soifer, Houck \& Neugebauer 1987), and therefore the dust in the galaxy must be
largely optically-thick to the starlight. However, in galaxies in the
standard optical catalogues, the ratio of far-infrared luminosity
to optical luminosity is often much less than one, suggesting that
the dust is largely optically thin to starlight (Soifer, Houck \&
Neugebauer 1987). It is, of course,
possible that even if much of the optical light is escaping from a galaxy,
the ultraviolet light, what the massive young stars mainly produce, 
may be
largely being absorbed by the dust, but, as far as we know, the quantitative
far-infrared---ultraviolet comparison of a large sample of galaxies which
is needed to answer this point has not been
done. The necessity of deciding whether galaxies are optically-thin
or optically-thick is because this determines how one predicts the
evolution of the submillimetre luminosity of a galaxy. If a galaxy is
optically-thick, the submillimetre luminosity is just proportional
to the star-formation rate in the galaxy, and the evolution of the
dust mass, and the elaborate apparatus we constructed \S2.1,
is irrelevant. If a galaxy is optically-thin, its submillimetre
luminosity depends on both the dust mass and the dust temperature, which
we can calculate from the star-formation rate. In reality, of course, a given
galaxy may pass through optically-thin and optically-thick stages. This
becomes too complex to model, and instead we have constructed some models
in which all the galaxies are assumed to be optically-thin and some
in which all the galaxies are assumed to be optically-thick.
\bigskip

\centerline{\bf 2.2.1. The optically-thin models}
\bigskip
If a galaxy is optically-thin to starlight, its submillimetre
luminosity, $L_{\nu}$, is given by
\medskip
$$
L_{\nu} = M_d \kappa_d(\nu) B(\nu, T_d),
$$
\medskip
\noindent in which $M_d$ is dust mass, $\kappa_d(\nu)$ is the dust absorption
per unit mass, and $B(\nu, T_d)$ is the Planck function. The absolute
value of $\kappa_d(\nu)$ is poorly known (Hughes et al. 1993) 
but its frequency dependence
is believed to lie between $\kappa_d \propto \nu$ 
and $\kappa_d \propto \nu^2$  (Whittet 1991). The temperature of the
dust, $T_d$, depends on the intensity of the interstellar radiation
field, and, if, as we assume, this is dominated by high-mass stars, 
then
$T_d \propto$ $\rm (star-formation\ rate)^{1/(4+n)}$, where $n$ is the index
in the relation $\kappa_d \propto \nu^n$. Thus, if we know the history
of star formation in a galaxy, we can calculate how both the dust mass and
the dust temperature depend on time. However, since we do not know
the constant of proportionality in the
relation between dust temperature and star-formation rate and since
the absolute value of the dust absorption coefficient is poorly
known (e.g. Hughes et al. 1993), we cannot immediately make an accurate prediction of how
the submillimetre luminosity depends on time. To do this, we have
to normalize the models to
the {\it observed} submillimetre properties of nearby galaxies.

Suppose there is a galaxy for which we know the 
far-infrared---submillimetre spectral energy distribution
and the relative amounts of gas and stars. We can estimate the
dust temperature from the spectral energy distribution, and 
if we make an assumption about the past star-formation rate
relative to the current star-formation rate, we 
can calculate how the dust temperature
has changed with time. Since we know the relative amounts of
gas and stars, we know the current value of the $f$ parameter (\S 2.1)
and hence we can calculate the evolution of the dust mass as a function
of the current dust mass. With this information, and since we know the
current submillimetre luminosity, we can calculate the evolution
of the submillimetre luminosity for this galaxy. Note that 
following this procedure means that we have avoided the
difficulty of not having an accurate value for the dust absorption
coefficient, and we also do not
need to know the current value of the star-formation
rate in the galaxy, something which observationally is very difficult
to measure. 

\bigskip
\centerline{\bf 2.2.2. The optically-thick models}

\bigskip

If a galaxy is optically-thick, the thermal emission from the dust in the
galaxy is simply proportional to the star-formation rate and is independent
of the evolution of the dust mass. Thus, if one knows the current
submillimetre luminosity of a galaxy and one makes an assumption
about its star-formation history, it is a simple matter to calculate
how its submillimetre luminosity should change with time. This skates
over one difficulty, because it is possible that the dust temperature will
also change, which, while keeping the total thermal emission the same, will
change the spectral energy distribution of the dust. However, whether or
not the dust temperature changes will depend on what assumptions one makes
about the distribution of the star formation, and the simplest first
assumption to make is that the dust temperature doesn't change.

\bigskip

\centerline{\bf 2.3. Statistical Predictions}

\bigskip

Once one has made assumptions about whether galaxies are optically-thin
or optically-thick and about the star-formation histories of galaxies,
it is straightforward to predict the submillimetre properties of the
high-redshift universe, as long as we know
the submillimetre luminosities 
of all the galaxies in the local universe (optically-thick case)
or the spectral energy distributions
and the
ratios of stellar to gas mass for all the galaxies in the local universe
(optically-thin case).
Of course we do not have this information, so we have to resort
to piecing together statistically the limited amount of information
that we do have.

The local submillimetre luminosity function for galaxies is not
known directly but, as a result of the IRAS survey,
the luminosity function at 60$\mu$m is known very well
(Lawrence et al. 1986). With some assumption about the 
average far-infrared---submillimetre
spectral energy distribution (SED) of galaxies it is therefore possible
to use the 60$\mu$m luminosity function to make an estimate of
the submillimetre luminosity function. The drawback of making
predictions based on this estimated local submillimetre
luminosity function
is that elliptical galaxies currently contain
very little dust and are therefore very weak sources of far-infrared
emission (Fich \& Hodge 1993), and are so not adequately 
represented in the 60$\mu$m
luminosity function. For this reason we have divided the galaxy population
into two classes: those well represented in the 60$\mu$m luminosity
function, a mixture of spirals and irregulars, which we will broadly label
as ``disk systems'', and elliptical galaxies. 

\bigskip

\centerline{\bf 2.3.1. Disk Galaxies}

\bigskip

Our method of predicting the contribution of disk galaxies to the
background and to the source counts is quite straightforward. We
start with the 60$\mu$m luminosity function of Lawrence et al.
(1986). Given the lack of submillimetre measurements of galaxies
(\S 1), it is impossible to do anything sophisticated such as investigating
whether a galaxy's SED in the far-infrared and
submillimetre wavebands is a function of morphology or far-infrared
luminosity. Instead, we have taken all the available data, both observations
of other galaxies and COBE results for our own galaxy, and shown that
all the observations are broadly consistent with a single SED
for disk galaxies (Fig. 2). This SED can be represented as
two grey-bodies with temperatures of 27K and 150K, with the ratio of the
masses of dust in the two components being $\rm
10^4$:1. For full details see the caption to Fig. 2. 

In the optically-thick case it is straightfoward to use the
standard cosmological formulae, this
SED, and the 60$\mu$m luminosity function to predict 
the submillimetre background and source counts. We have
constructed two basic models. In the first, we just assume that the
star-formation rate does not change with time, something
which is suggested by optical observations of
nearby spiral galaxies (Kennicutt 1983). In this case
the submillimetre luminosity of a galaxy will also be a constant.
In
the second case we used the
star-formation history inferred by Pei and Fall (1995) from a study of the
statistics of Ly$\alpha$ absorption systems. In this the star-formation rate
increases rapidly from the current epoch to a redshift of one, falling off
thereafter; the initial rapid rise in star-formation rate being similar
to that inferred from the Canada-France Redshift Survey (Lilly et al. 1996).
As we discussed above, in the optically-thick approximation the
submillimetre luminosity simply scales as this star-formation
rate.
In both models we only considered galaxies out to 
a maximum redshift ($\rm z_{max}$), equivalent to assuming that the galaxies 
form at this
redshift. The predicted backgrounds for $\rm z_{max} = 3$ are shown in
Figure 3 and those for $\rm z_{max} = 1$ in Figure 4. The background
that Puget et al. (1996) have claimed to find in the COBE database
is shown in both figures, as are the upper limits on the background
that Davies et al. (1996) have derived from the DIRBE experiment on
COBE. Figure 6 shows the predicted source counts at 190$\mu$m, one of
the main ISO wavelengths, and Figure 7 shows the predicted source
counts at $\rm 850\ \mu m$, the main SCUBA wavelength. 

In the optically-thin case, we have to make an additional assumption
about the ratio of gas mass to stellar mass in present-day galaxies.
Following the results of Young \& Scoville (1991), we assume that
all present-day disk galaxies
have a ratio of gas mass to stellar
mass of 1:9 (U\footnote\dag{This is undoubtedly 
simplistic, because Young and Scoville's
result show that this ratio is a function of Hubble type, but it is
the best we can do at the moment. 
If present-day
disk galaxies generally have a smaller ratio
of gas to dust, the submillimetre luminosities at high redshift
will be increased (Fig. 1), and vice versa.}). 
We tried 
the same two
star-formation histories as in the optically-thin case and
used a number of different models for the evolution of the dust mass.
We also tried the same two values for $\rm z_{max}$ as were used in the
optically-thick case.
The predicted backgrounds are also shown in Figs 3 \& 4, where we have tried to
show enough models that the effect of adjusting the different
input parameters ($\rm \Omega_0$, changing from inflow to outflow, changing
the dust parameter in an outflow model, to give three examples) is
clear. Figures 6 \& 7 show the predicted source counts. 
\bigskip

\centerline{\bf 2.3.2. Elliptical Galaxies}
\bigskip

Our method of predicting the contribution of elliptical galaxies
to the background and the source counts is more
complicated and, in order not to complicate the method even more,
we just consider optically-thin models. We consider the qualitative
effect of changing from an optically-thin model to an optically-thick
model at the end of this section. 

We assume that star
formation in a particular galaxy starts at some redshift, continues at a
constant rate for
a fixed time, and then stops, all the gas having been used up---a process
we loosely refer to as `galaxy formation'. 
The parameters
that have to be put into the model are the range of redshifts over which
galaxy formation occurs and the time it takes an individual galaxy to
form. If this time, $\tau$, is less than the duration of the galaxy-formation
epoch, $\tau_{gf}$, we assume that the fraction of galaxies that are
forming at any one time during this epoch is $\tau / \tau_{gf}$. 

Before we
continue with the details of the model it is instructive to consider some
of its broad implications. During the formation period the dust
mass in the galaxy follows one of the curves shown in Fig. 1, ending
at zero as we are assuming that the formation period ends when all the
gas has been turned into stars. Changing the
value of $\tau$ will not change the maximum dust mass, it will merely cause
the dust mass to continue along its evolutionary track either more or less
quickly. However, the submillimetre luminosity of the galaxy will change
because the stars are forming at a different rate, and thus the interstellar
radiation field and hence the dust temperature will be different. Changing
the value of $\tau$ from 1 Gyr to 0.1 Gyr, for example, will increase the star
formation rate by a factor of 10, and hence the dust temperature by a factor
of $10^{1/(4+n)} \sim 1.5$. The effect of this rise in dust temperature
on the predicted submillimetre flux will depend greatly on the wavelength
of the observation. If one is observing on the Rayleigh-Jeans side of the
thermal peak, the increase in submillimetre flux will be only this same
modest factor, whereas if one is observing on the Wien side the increase can be
very large indeed. This is an example of the general point that observations
on the Rayleigh-Jeans side of the thermal peak are primarily sensitive
to dust mass, whereas those on the Wien side are primarily sensitive to
dust temperature. This argument can also be used to place an approximate upper
limit on the dust temperature of an elliptical galaxy. Disk galaxies typically
have dust temperatures of 20-30K (Eales, Wynn-Williams \& Duncan 1989;
Stark et al. 1989). If we assume that these have been forming stars at
roughly the same rate for the age of the universe, and if we take the free-fall
time ($\simeq$0.1 Gyr---Fall \& Rees 1985) as the smallest 
value for the formation period
of elliptical galaxies, the maximum dust temperature for elliptical 
galaxies with the same mass as low-redshift 
spirals is $\sim 25 \times 1000^{1/(4+n)} \simeq 88$K, consistent with
the dust temperatures that have been measured for two high-redshift objects
(Downes et al. 1992; Isaak et al. 1994).

The problem in implementing the models, as with the disk galaxies, is that
we need some way of normalizing the models to  the observed submillimetre
properties of galaxies, in order to avoid the necessity of having to calculate
absolute dust masses and of needing to know the constant of proportionality
in the relation between dust temperature and star-formation rate. We can
do this for elliptical galaxies, but in a less satisfactory way than for
disk galaxies. We use the spiral galaxy NGC 4254, which was observed
by Stark et al. (1989) and has a dust temperature (assuming $\kappa_d \propto 
\nu^2$)
of 23K. This is not a special galaxy, and any galaxy would do for which
it is possible to estimate a dust temperature. We assume that, as for most
spiral galaxies (Young and Scoville 1991), the ratio of gas mass to 
total mass (gas and stars) is 0.1. Once we have done this, we can
calculate the submillimetre luminosity of an elliptical galaxy by scaling
from the properties of NGC 4254. The submillimetre luminosity
of the elliptical galaxy is given by
\bigskip
$$
L_{\nu} = L_{\nu, NGC 4254} 10^{(M_{NGC 4254} - M) \over 2.5} 
{(M/L)_{elliptical} \over (M/L)_{disk}} \times 23 ({\tau_h \over \tau})^{1/(4+n)}
\times ({M_d(f) \over M_d(0.1)}),
$$
\bigskip
\noindent in which 
$L_{\nu, NGC 4254}$ is the submillimetre luminosity of NGC 4254, $M$ is
the absolute magnitude the elliptical galaxy has at the current epoch,
$M_{NGC 4254}$ is the absolute magnitude of NGC 4254, $\tau_h$ is the
age of the universe, and $M_d$ is the dust mass as a function of the
parameter $f$, which measures how far the elliptical galaxy is through the
formation process (\S 2.1). Using the results of
Persic and Salucci (1992), we assume that the ratio of the 
baryonic mass-to-light of ellipticals to that of spirals is 3.6. 
We
use a value for $n$ of
2.

Figure 5 shows the predicted background for a number of models. We
have again tried to
show enough models that the effect of adjusting the different
input parameters is clear. Figures 8 and 9 show the source counts
predicted at 190 and 850 $\mu$m. 

In order not the increase the complexity of the models, we have not
constructed optically-thick models. It is clear, however, that the
qualitative 
effect on the background predicted by a particular model of going
from the optically-thin assumption to the optically-thick assumption
would be 
to increase the brightness of the background, since none of the optical
light would then be escaping, but to leave the spectral shape roughly
the same, since this is largely dependent on the redshift at which the
galaxies are assumed to form.

\bigskip
\centerline{\bf 3. Discussion}
\bigskip

One of the most obvious features of the disk-galaxy predictions
is that at short wavelengths ($\lambda < 200 \mu m$) all of the
models predict a higher background than was actually measured by Puget
et al. (1996).
There are two obvious possible explanations of this.
First,
Puget et al.'s estimate of the background may be too low,
which is perfectly
possible given the complicated technique they were forced to
use to remove the foreground emission.
Second, our models may be wrong in some way.
This might be caused by a fundamental error in the input physics, 
or it might be caused by one of
the simplifying assumptions we were forced to make because of our
ignorance of the
submillimetre properties of the nearby
universe, on which the models rely. We can think of two main problems.
First, we do not have a direct measure of the local submillimetre luminosity
function and were forced to extrapolate from the far infrared.
Second, we had
to ``lump'' all disk galaxies together and
and assume a single far-infrared---submillimetre spectral energy 
distribution
and a single value for the ratio of gas mass to stellar mass; 
and we already know this
second assumption, at least, is wrong, because Young \& Scoville
(1991)
have found that this ratio is a function of Hubble type. 
The advent of SCUBA should remove most of these problems,
because with SCUBA it will be possible to measure the submillimetre
luminosity function directly, and once
submillimetre fluxes
exist for large numbers of local galaxies it will be possible 
to predict the background produced by different groups of galaxies, divided
either by Hubble type,
far-infrared luminosity, or by some other criterion.
 
Although all the models predict too much background at the short
wavelengths, there are two models which predict a background which
is much greater than Puget et al.'s background estimate. These
models also exceed the upper limit on the background estimated
by Davies et al. (1996) from COBE data, and since these limits were
produced in a relatively straightforward way, we feel that here the
problem must clearly lie with the models. The two models that are
very discrepant are the ones in which we assume the star-formation
history proposed by Pei \& Fall (1995) from a study of the statistics
of quasar absorption lines, a history in which
the star-formation rate at $\rm z \sim 1$ is much greater
than that today, something which has also been claimed
on the basis of the results from
the Canada-France Redshift Survey (Lilly et al. 1996). Since one
of the two models is the optically-thin model and the other the
optically-thick model, it seems clear that if the star-formation
rate in galaxies is increasing rapidly with redshift, we are not seeing
any evidence for this in the submillimetre waveband. 

At longer wavelengths ($\lambda > 350\ \mu m$), the disk-galaxy 
predictions are generally lower than
the observed background, but still make a significant contribution to
the background.

The elliptical galaxy models lead to a wide variety of predictions, and it
is clear from an inspection of Fig. 5 that a fine-tuning of the parameters
would lead to a model that would fit the background rather well. The model
that produces a background with a shape most similar to that of the observed
background is a closed-box model with galaxies forming during the range of
redshifts $\rm 2 < z < 5$, with a formation period of 1 Gyr. The predicted
background is about a factor of two lower than the observed background, but
this is not a large factor given the extra uncertainties that went into
the elliptical models (\S 2.3.2). Nevertheless, we caution
that simply trying to match the elliptical models to the observed
background is not appropriate, since the disk-galaxy models show that
a significant fraction of the background must be produced by galactic
disks.

One surprising feature of our results is that the disk galaxies appear to
be producing a stronger background than the elliptical galaxies, which is
in conflict with arguments based on global metallicity. The integrated
background light produced by a population
of objects at a redshift $z$ is related to the smoothed-out
cosmic density of processed material, $<\rho (Z + \Delta Y)> $, produced by those objects by
\bigskip
$$
\int^{\infty}_0 I_{\nu} d\nu = {0.007 <\rho (Z + \Delta Y >c^3 \over 4 \pi (1 + z)}
$$
\bigskip
\noindent (e.g. Pagel 1993). Pagel claims
that the smoothed-out density of
metals produced by elliptical
systems is $\simeq$6 times greater than that 
produced by spirals, which should lead, using this
equation and ignoring for the moment the redshift factor, 
to a background from the
elliptical galaxies $\simeq$6 times higher than that from 
the spirals. We find rather
the reverse, with there being typically (the exact difference depends on which
models are being compared) a factor of five the other 
way. We can account for part
of this difference with 
the redshift factor. In our elliptical models the galaxies
are at much higher redshifts than in 
the disk models. However, this usually only
amounts to a factor of $\sim 3-4$, still leaving a 
factor of $\sim$10 discrepancy.

The only explanation we can think of for this discrepancy is if much of the
far-infrared---submillimetre emission from spiral galaxies is not from massive
but from low-mass stars---in which case the equation above is invalid. 
As we discussed in \S 2.2
there is a long-running
argument in the literature as to whether 
the bulk of the far-infrared emission from galaxies
is from dust heated by old low-mass stars or from high-mass stars.
In our models we have assumed the latter, but if the former is true
this would at least partly explain the discrepancy, especially 
because the
contribution of the high-mass stars, which will generally produce hotter
dust, will be less in the submillimetre waveband. It would also, of 
course,
invalidate the relationship between dust temperature and 
star-formation rate that we have
assumed in our optically-thin models. However, this is probably not 
too important. In 
most of the
disk models we have assumed a constant star-formation rate, so this would not
be a concern, and in models in which 
the star-formation rate increases with redshift, although
the relation may not be true at the lowest redshifts, it will eventually become true once the
heating effect of the massive stars dominates the 
heating effect of the low-mass stars.
For the elliptical models this should 
not be a concern, as the star-formation rates, and hence
the heating effects of the massive stars, will 
be much greater than the star-formation rates
in low-redshift spiral galaxies. It is also a potential problem for
our two optically-thick disk-galaxy models, in which we assume that 
the submillimetre
luminosity is proportional to the star-formation rate. However,
in the model in which the star-formation
rate is a constant this is not a problem, because the luminosity
function at every redshift is exactly the same as the local luminosity
function, which necessarily takes into account all the stars that are
heating the dust. It is also probably not a major problem for the
optically-thick model with the evolving star-formation rate, since
although submillimetre luminosity may not be proportional to the
star-formation rate at low redshift, it may well be true at high redshifts
where the star-formation rate is higher.

Among the wide variety of source counts predicted by the different models (Figs 6-9) we can pick out
one interesting qualitative feature. The 850 $\mu$m counts for the different elliptical
galaxy models all have the same shape, although they are offset horizontally and
vertically from each other,
except for the counts predicted using the one model in which we changed the shape
of the low-redshift optical luminosity function. The reason for this is due
to the characteristic spectral energy distribution of thermal dust emission.
It is known (e.g. Hughes 1996) that if 
one is observing a galaxy at a wavelength well on the long-wavelength side of the
peak of the thermal dust emission, the galaxy's flux will be approximately independent of redshift
from $\rm z \sim 1$ to the redshift at which emitted wavelength gets close to the
wavelength of the peak. The reason for this is that at $\rm z > 1$ the effect of 
increasing luminosity-distance on the flux is
almost exactly cancelled by the fact that the emitted wavelength
gets closer to the
wavelength of the thermal peak, until the emitted wavelength is at the peak, at which redshift the
two effects start to act in the same direction.
At an observing wavelength of 850 $\mu$m, the 
high temperature expected for the dust in spheroids in their formation phase means that this cancellation
occurs from $\rm 1 \leq z \leq 10$. Thus, in this redshift range, which is where most of the
sources are, there is a one-to-one mapping between submillimetre
flux and luminosity, and the shape of the source counts will be the same 
as the shape of the distribution of submillimetre luminosities. Since the submillimetre
luminosity at long wavelength is much more 
sensitive to dust mass than dust temperature (\S2.3.2),
this shape will also be the shape of the distribution of dust masses, which in turn will be
approximately the same as the shape of the distribution of total masses; hence the sensitivity
of the shape of the counts to the shape of the input luminosity function. Thus the shape of the counts
is mainly governed by the distribution of masses of the systems that are doing the radiating,
whereas the position of the counts in the diagram is governed by a combination of all 
the other factors that go into the models.

Finally, we look to the future. SCUBA and ISO should solve many of the
problems with the disk-galaxy models: the reliance on a far-infrared
luminosity function, rather than on a direct measurement of the submillimetre
luminosity function; the gross simplification that all local galaxies have
the same ratio of stellar mass to gas mass and the 
same far-infrared---submillimetre luminosity function. One additional
limitation of our models is that we have made no attempt
to incorporate the
effects of different
types of structure evolution, 
a choice we made partly because there is no consensus about how it 
occurs and partly because there is already sufficient 
variety in the predictions when we
include more well-understood types of evolution. 
However, in principle there is no difficulty adapting the models
to include it, and one can already use arguments based on the present
models to show that including hierarchical clustering
should have a relatively small 
effect on the background\footnote\dag{In the hierarchical
clustering scenario, one of the high-redshift galaxies 
in our models will actually be divided into
a number of smaller units. However, the same mass of dust and same 
number of stars will be present.
If we consider the optically-thin case first,
the intensity of the interstellar radiation field in these smaller 
units may be somewhat different because
of the different geometry, but the dust 
temperature depends only weakly on the
intensity of the interstellar radiation field, and thus the 
total contribution to the background
will be quite similar to what it is in the case of a single large galaxy. 
In the optically-thick case, all the starlight is assumed to be absorbed
by dust, no matter the size of the structure in which the dust and stars
reside, and so the predicted background will not change.}.
Finally SCUBA and ISO will allow us to test both the fundamental physics
on which the models are based---once dust has been produced, does
it stay within the galaxy?---and the models directly by producing observed
source counts with which to confront the predictions.
\bigskip

\centerline{\bf Acknowledgements}
\bigskip

The first draft of this paper was 
written while SAE was a guest
at the Canadian Institute for Theoretical Astrophysics.
He thanks the staff there for their hospitality.
We thank Mike Fall and Gerry Gilmore for their comments on the first
version of this paper.

\bigskip
\item{} Babul, A. \& Rees, M.J. 1992, MNRAS, 255, 346.
\smallskip
\item{} Beichman, C.A. \& Helou, G. 1991, ApJ, 370, L1.
\smallskip
\item{} Blain, A.W. \& Longair, M.S. 1993, MNRAS, 264, 509.
\smallskip
\item{} Blain, A.W. \& Longair, M.S. 1996, MNRAS, 279, 847.
\smallskip
\item{} Bond, J.R., Carr, B.J. \& Hogan, C.J. 1986, ApJ, 306, 428.
\smallskip
\item{} Boulanger, F. \& Perault, M. 1988, ApJ, 330, 964.
\smallskip
\item{} Davies, J.I. et al. 1996, MNRAS, submitted.
\smallskip
\item{} De Propris, R., Pritchett, C.J., Hartwick, F.D.A. \&
Hickson, P. 1993, AJ, 105, 1243.
\smallskip
\item{} Devereux, N. \& Young, J.S. 1990, ApJ, 350, L25.
\smallskip
\item{} Downes, D., Radford, S.J.E., Greve, A., Thum, C., 
Solomon, P.M. \& Wink, J.E. 1992, ApJ, 398, L25.
\smallskip
\item{} Duncan, W.D. 1990, {\it Submillimetre Astronomy}, eds
Watt, G.D. \& Webster, A.S., Kluwer (Dordrecht), p51.
\smallskip
\item{} Eales, S.A. 1991, {\it After the First Three Minutes}, eds
S. Holt, C. Bennett \& V. Trimble (AIP), p206.
\smallskip
\item{} Eales, S.A. \& Edmunds, M.G. 1996, MNRAS, 280, 1167.
\smallskip
\item{} Eales, S.A., Wynn-Williams, C.G. \& Duncan, W.D. 1989,
ApJ, 339, 859.
\smallskip
\item{} Edmunds, M.G. 1990, MNRAS, 246, 678.
\smallskip
\item{} Fall, S.M., Charlot, S. \& Pei, Y.C. 1996, ApJ, 464, L43.
\smallskip
\item{} Fall, S.M. \& Rees, M.J. 1985, ApJ, 298, 18.
\smallskip
\item{} Fich, M. \& Hodge, P. 1993, ApJ, 415, 75.
\smallskip
\item{} Franceschini, A., Mazzei, P., De Zotti, G. \& 
Danese, L. 1994, ApJ, 427, 140.
\smallskip
\item{} Helou, G. 1986, ApJ, 311, L33.
\smallskip
\item{} Hildebrand, R.H. 1983, QJRAS, 24, 267.
\smallskip
\item{} Hughes, D.H., Robson, E.I., Dunlop, J.S. \& Gear, W.K. 1993, MNRAS, 263, 607.
\smallskip
\item{} Hughes, D.H. 1996, {\it Cold Gas at High Redshift}, ed. M. Bremer, Kluwer, in press.
\smallskip
\item{} Isaak, K.G., McMahon, R.G., Hills, R.E. \& Withington, S.
1994, MNRAS, 267, L9.
\smallskip
\item{} Kennicutt, R.C. 1983, ApJ, 272, 54.
\smallskip
\item{} Koo, D.C., Guzm\'an, R., Faber, S.M., Illingworth, G.D., Bershady, M.A., Kron, R.G.
\& Takamiya, M. 1995, ApJ, 440, L49.
\smallskip
\item{} Lawrence, A., Walker, D., Rowan-Robinson, M., Leech, K.J. \&
Penston, M.V. 1986, MNRAS, 219, 687.
\smallskip
\item{} Lilly, S.J., Le F\`evre, O., Hammer, F. \& Crampton, D. 1996, ApJ,
460, L1.
\smallskip
\item{} Pagel, B. 1993, preprint. 
\smallskip
\item{} Pei, Y.C. \& Fall, S.M. 1995, ApJ, 454, 69.
\smallskip
\item{} Persic, M. \& Salucci, P. 1992, MNRAS, 258, 14p.
\smallskip
\item{} Persson, C.J. \& Helou, G. 1987, ApJ, 314, 513.
\smallskip
\item{} Puget, J.-L., Abergel, A., Bernard, J.-P., Boulanger, F.,
Burton, W.B., Desert, F.-X. \& Hartmann, D. 1996, 
A \& A, 308, L5.
\smallskip
\item{} Schade, D., Lilly, S.J., Crampton, D., Hammer, F., Le F\`evre, O. \& Tresse, L. 1995,
ApJ, 451, L1.
\smallskip
\item{} Soifer, B.T., Houck, J.R. \& Neugebauer, G. 1987, ARAA, 25, 187.
\smallskip
\item{} Stark, A.A., Davidson, J.A., Harper, D.A., Pernic, R.,
Loewenstein, R., Platt, S., Engargiola, G. \& Casey, S. 1989, ApJ, 337, 650.
\smallskip
\item{} Wang, B. 1991a, ApJ, 374, 456.
\smallskip
\item{} Wang, B. 1991b, ApJ, 374, 465.
\smallskip
\item{} Whittet, D.C.B. 1991, {\it Dust in the Galactic Environment},
Institute of Physics Publishing.
\smallskip
\item{} Wright, E.L. et al. 1991, ApJ, 381, 200. 
\smallskip
\item{} Young, J.S. \& Scoville, N.Z. 1991, ARAA, 29, 581.
\vfill
\eject
\centerline{\bf Figure Captions}
\parindent = 0pt
\bigskip
Fig. 1: Elementary models for the evolution of the mass of dust in a galaxy.
The ordinate $R$ is the ratio of mass of interstellar dust at gas fraction
$f$ to the interstellar dust mass at $f=0.1$. The continuous
curve is for the
simple closed-box model. The dashed curves are for outflow models, with
the lowest one having
$\lambda/\alpha = $1.0, 
the middle one having 
$\lambda/\alpha = $3.0, 
and the highest one having
$\lambda/\alpha = $10.0. 
The dot-dashed curves are for
inflow
models with the lower having $\gamma/\alpha = $0.99, and the
higher having $\gamma/\alpha = $0.9.
\bigskip
Fig. 2: A synthesis of the submillimetre observations that exist 
for nearby galaxies. For constructing a spectral energy distribution
between the far-infrared and the submillimetre wavebands the most useful
measurements that exist are the COBE spectrum of our galaxy (Wright et al.
1991) and the submillimetre measurements of other
galaxies by Eales, Wynn-Williams \& Duncan
(1989) and by Stark et al. (1989), more recent measurements being made
through a smaller aperture which exacerbates the problem of matching
the submillimetre flux measurements to the IRAS far-infrared measurements.
Despite the large aperture used by Eales et al., there is a problem correcting
the IRAS measurements to the same aperture (Eales et al. 1989), whereas 
this is not a problem
for the Stark et al. measurements because they measured integrated fluxes.
For this reason we used the Stark et al. data to investigate the shape
of the spectral energy distribution at wavelengths less than 360$\mu$m,
their longest-wavelength measurement, using the Eales et al. data at
longer wavelengths. The figure shows the Stark et al. fluxes normalized
so that each galaxy has a flux of 100 Jy at 100 $\mu$m, with the Eales et al.
fluxes at $\lambda \geq 350\mu$m normalized to a 350 $\mu$m flux of 22.5 Jy,
a figure chosen so that the shorter wavelength measurements would approximately
agree with the Stark et al. data at $\lambda \sim 350\mu$m. Each galaxy is 
represented by a different symbol, and the spread of the symbols at the
different wavelengths is an indication of the variation in galaxy
SED's. The COBE spectrum of the galaxy has also been plotted on the
diagram (thick line) and has been normalized to give as good an agreement as possible
with the fluxes of the galaxies. The agreement between the SED's of the
different galaxies is remarkably good (especially if one considers that
some of dispersion is caused by measurement errors), and thus, given the
present data, it is reasonable to use a single SED in 
the cosmological modelling of the disk population. The thin line shows
the analytic model that is a good fit to the data and is used in the
modelling. It consists of thermal emission from
two dust components: one at 27K (dash) and one at 
135K (dot-dash), with the mass of cold dust being $\rm 10^4$ times greater
than the mass of hot dust. We have assumed the emissivity law suggested
by Hildebrand (1983), in which $\kappa_d \propto \nu^2$ at 
$\lambda > 250 \mu$m and $\kappa_d \propto \nu$ at $\lambda < 250 \mu$m. 
\bigskip
Figure 3: Predicted backgrounds from disk galaxies. The thick dot-dash
line shows the background detected by Puget et al. (1996), the upper limits
are the
upper limits on the background determined by Davies et al. (1996).
The thin lines show the predictions of various models. In all of the
models disk galaxies are assumed to form at $\rm z = 3$. 
Unless otherwise stated, 
it is assumed that $\rm \Omega_0 = 1$ and that the star-formation rate
is a constant.
Lines A to F are the predictions of the optically-thin models. Line A
(solid line) is the prediction of the closed-box model. Lines
B-D (dashed) are the predictions for inflow and outflow models
(B---inflow, $\gamma/\alpha = 0.99$; C---outflow, $\rm \lambda/\alpha = 1.0$;
D---outflow, $\rm \lambda/\alpha = 10.0$). Line E (dot-dash) is for
a closed-box model again, but with the        
star-formation history derived by Pei and Fall (1995; see text).
Line F (dots) is for a closed-box model, but with $\rm \Omega_0 = 0$.
Lines G and H (dot-dot-dot-dash) are for the optically-thick models, line
G being the prediction for the model in which the star-formation
rate is a constant, and line H for the model with the Pei and Fall 
star-formation history.
\bigskip

Figure 4: Predicted backgrounds from disk galaxies. The thick dot-dash
line shows the background detected by Puget et al. (1996), the upper limits
are the
upper limits on the background determined by Davies et al. (1996).
The thin lines show the predictions of various models. In all of the
models disk galaxies are assumed to form at $\rm z = 1$.
Unless otherwise stated,
it is assumed that $\rm \Omega_0 = 1$ and that the star-formation rate
is a constant.
Lines A to F are the predictions of the optically-thin models. Line A
(solid line) is the prediction of the closed-box model. Lines
B-D (dashed) are the predictions for inflow and outflow models
(B---inflow, $\gamma/\alpha = 0.99$; C---outflow, $\rm \lambda/\alpha = 1.0$;   
D---outflow, $\rm \lambda/\alpha = 10.0$). Line E (dot-dash) is for    
a closed-box model again, but with the 
star-formation history derived by Pei and Fall (1995; see text).
Line F (dots) is for a closed-box model, but with $\rm \Omega_0 = 0$.
Lines G and H (dot-dot-dot-dash) are for the optically-thick models, line
G being the prediction for the model in which the star-formation
rate is a constant, and line H for the model with the Pei and Fall 
star-formation history.
\bigskip

Figure 5: Predicted backgrounds from elliptical galaxies. The thick dot-dash
line shows the background detected by Puget et al. (1996), the upper
limits are the
upper limits on the background determined by Davies et al. (1996).
The thin lines show the predictions of various models. In all of the
models but one we assume that an individual galaxy forms in 0.1 Gyr, and
in all the models but one we assume that $\rm \Omega_0 = 1$.
The solid lines are for a closed-box model. Line A is for
a model
in which the galaxies form in the redshift interval $\rm 5 < z < 10$; line
B is for 
a model in which the galaxies form in the redshift
range $\rm 2 < z < 5$ with an individual galaxy forming in 1 Gyr; line C is
for a model in which 
the galaxies form in the redshift
range $\rm 2 < z < 5$ with an individual galaxy forming in 0.1 Gyr.
The dotted line (D) is for an inflow model ($\rm \gamma/\alpha = 0.9$)
in which galaxies form in the redshift range $\rm 5 < z < 10$. The
dot-dash line (E) is for a closed-box model again, in which the galaxies
form in the redshift range $\rm 5 < z < 10$ but in which $\rm \Omega_0 = 0$.
The dashed line (F) is for an outflow model ($\lambda/\alpha = 4.0$) in
which galaxies form in the redshift range $\rm 5 < z < 10$.
\bigskip
Figure 6: Differential source counts at 190 $\mu$m for disk galaxy models.
The ordinate is number arcmin$^{-2}$. 
In all of the
models disk galaxies are assumed to form at $\rm z = 3$.
Unless otherwise stated, we assume $\rm \Omega_0 = 0$ and that
the star-formation rate does not change with time.
The symbols are as follows: optically-thin closed-box---circles with
dots in centres; optically-thin inflow ($\rm \gamma/\alpha =0.99$)---dots;
optically-thin outflow ($\rm
\lambda/\alpha = 1.0$)---circles;
optically-thin outflow ($\rm
\lambda/\alpha = 10.0$)---crosses;
optically-thin closed-box with star-formation history
from Pei \& Fall (1995, see text)---squares;
optically-thin closed-box with $\rm \Omega_0 = 0$---triangles;
optically-thick---crosses in circles; optically-thick with Pei
and Fall star-formation history---asterisks.

\bigskip
Figure 7: Differential source counts at 850 $\mu$m for disk galaxy models.
The ordinate is number arcmin$^{-2}$.
In all of the
models disk galaxies are assumed to form at $\rm z = 3$.
Unless otherwise stated, we assume $\rm \Omega_0 = 0$ and that
the star-formation rate does not change with time.
The symbols are as follows: optically-thin closed-box---circles with
dots in centres; optically-thin inflow ($\rm \gamma/\alpha =0.99$)---dots;
optically-thin outflow ($\rm 
\lambda/\alpha = 1.0$)---circles;
optically-thin outflow ($\rm 
\lambda/\alpha = 10.0$)---crosses;  
optically-thin closed-box with star-formation history
from Pei \& Fall (1995, see text)---squares;
optically-thin closed-box with $\rm \Omega_0 = 0$---triangles;
optically-thick---crosses in circles; optically-thick with Pei
and Fall star-formation history---asterisks.
\bigskip
Figure 8: Differential source counts at 190 $\mu$m for elliptical galaxy models.
The ordinate is number arcmin$^{-2}$.
In all the models but one we assume that $\rm \Omega_0 = 1$ and we assume a closed-box
model for dust evolution unless otherwise stated. In all the models but one we have assumed
the optical luminosity function mentioned in the text. In the one exception we have investigated the
effect of changing the luminosity function, by changing the value of the Schechter $\alpha$
parameter to -1.0, while leaving the other Schechter parameters the same. In this model, for
which the predictions are shown by open circles with crosses inside, we have assumed that
galaxies form in the redshift range $\rm 5 < z_f < 10$ and that an individual galaxy forms
in a period ($\tau$) of 0.1 Gyr. The symbols for the other models are as follows:
$\rm 2 < z_f < 5$, $\rm \tau =$ 1 Gyr---dots in circles;
$\rm 2 < z_f < 5$, $\rm \tau =$ 0.1 Gyr---dots;
$\rm 5 < z_f < 10$, $\rm \tau =$ 0.1 Gyr---open circles;
outflow ($\rm \lambda/\alpha = 4.0$), $\rm 5 < z_f < 10$, $\rm \tau =$ 0.1 Gyr---diagonal crosses;
inflow ($\rm \gamma/\alpha = 0.9$), $\rm 5 < z_f < 10$, $\rm \tau =$ 0.1 Gyr---squares;
$\rm \Omega_0 = 0$, $\rm 5 < z_f < 10$, $\rm \tau =$ 0.1 Gyr---triangles.
\bigskip

Figure 9: Differential source counts at 850 $\mu$m for elliptical galaxy models.
The ordinate is number arcmin$^{-2}$.
In all the models but one we assume that $\rm \Omega_0 = 1$ and we assume a closed-box
model for dust evolution unless otherwise stated. In all the models but one we have assumed
the optical luminosity function mentioned in the text. In the one exception we have investigated the
effect of changing the luminosity function, by changing the value of the Schechter $\alpha$
parameter to -1.0, while leaving the other Schechter parameters the same. In this model, for
which the predictions are shown by open circles with crosses inside, we have assumed that
galaxies form in the redshift range $\rm 5 < z_f < 10$ and that an individual galaxy forms
in a period ($\tau$) of 0.1 Gyr. The symbols for the other models are as follows:
$\rm 2 < z_f < 5$, $\rm \tau =$ 1 Gyr---dots in circles;
$\rm 2 < z_f < 5$, $\rm \tau =$ 0.1 Gyr---dots;
$\rm 5 < z_f < 10$, $\rm \tau =$ 0.1 Gyr---open circles;
outflow ($\rm \lambda/\alpha = 4.0$), $\rm 5 < z_f < 10$, $\rm \tau =$ 0.1 Gyr---diagonal crosses;
inflow ($\rm \gamma/\alpha = 0.9$), $\rm 5 < z_f < 10$, $\rm \tau =$ 0.1 Gyr---squares;
$\rm \Omega_0 = 0$, $\rm 5 < z_f < 10$, $\rm \tau =$ 0.1 Gyr---triangles.
\vfill
\supereject
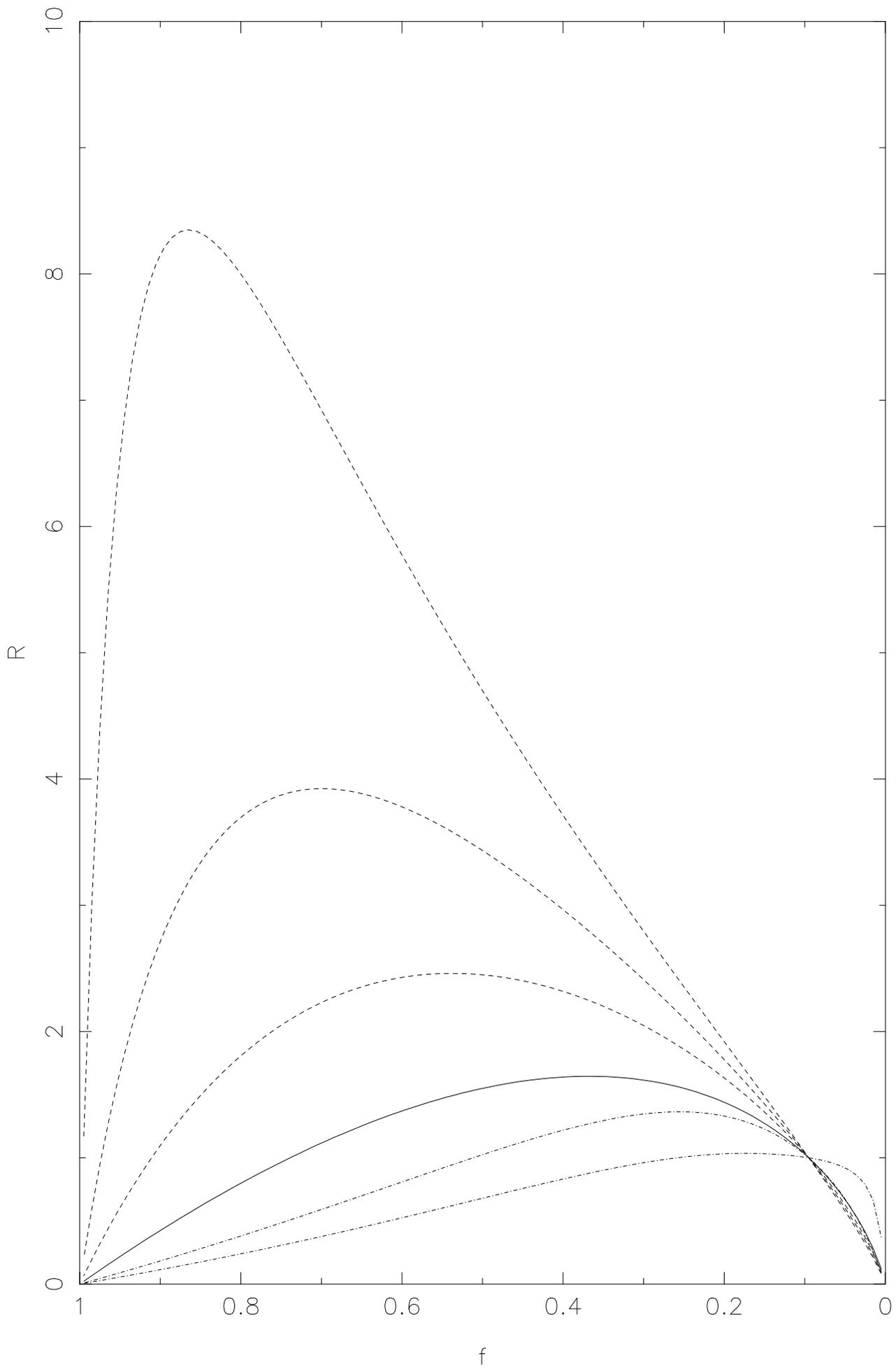
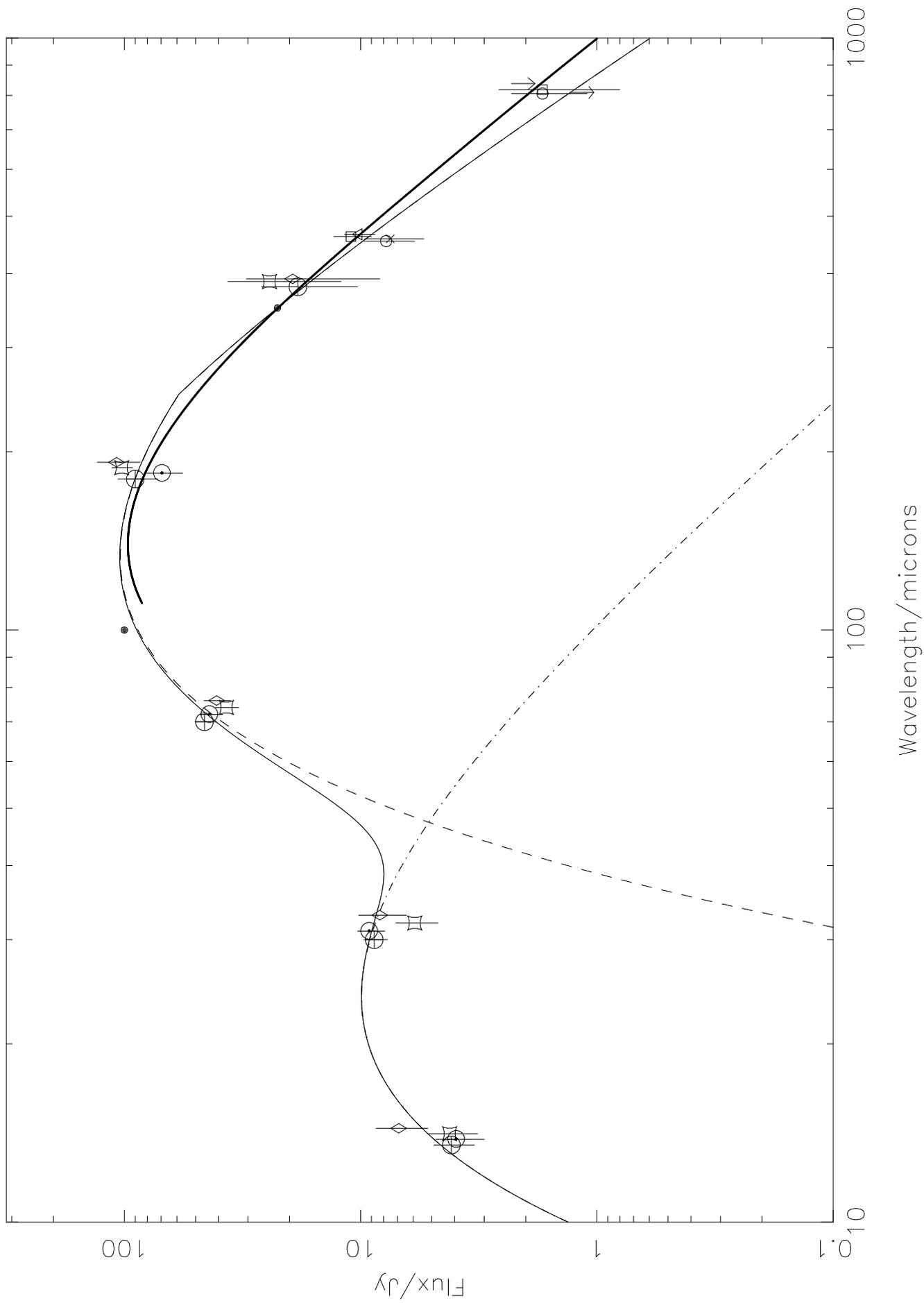
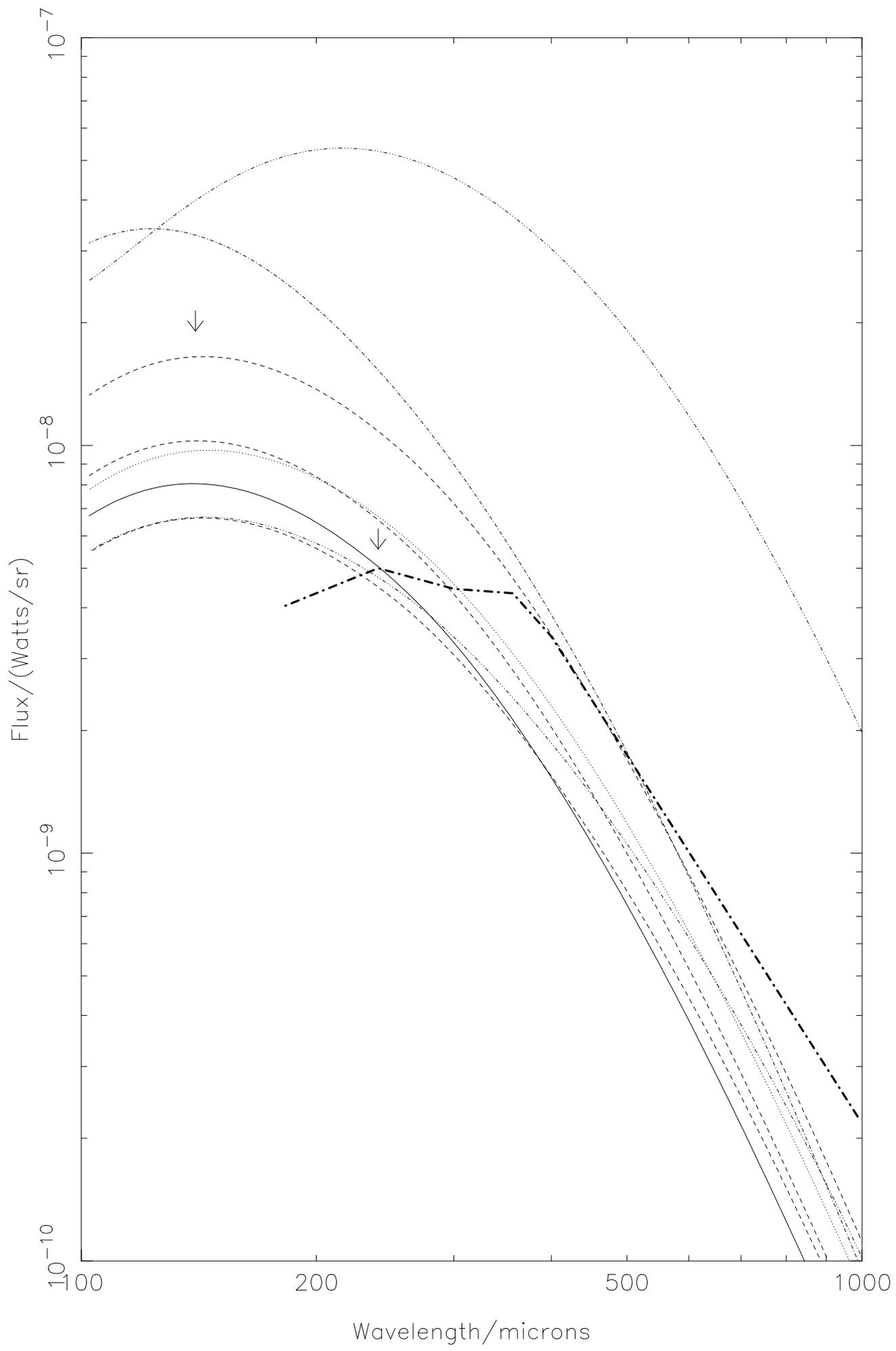
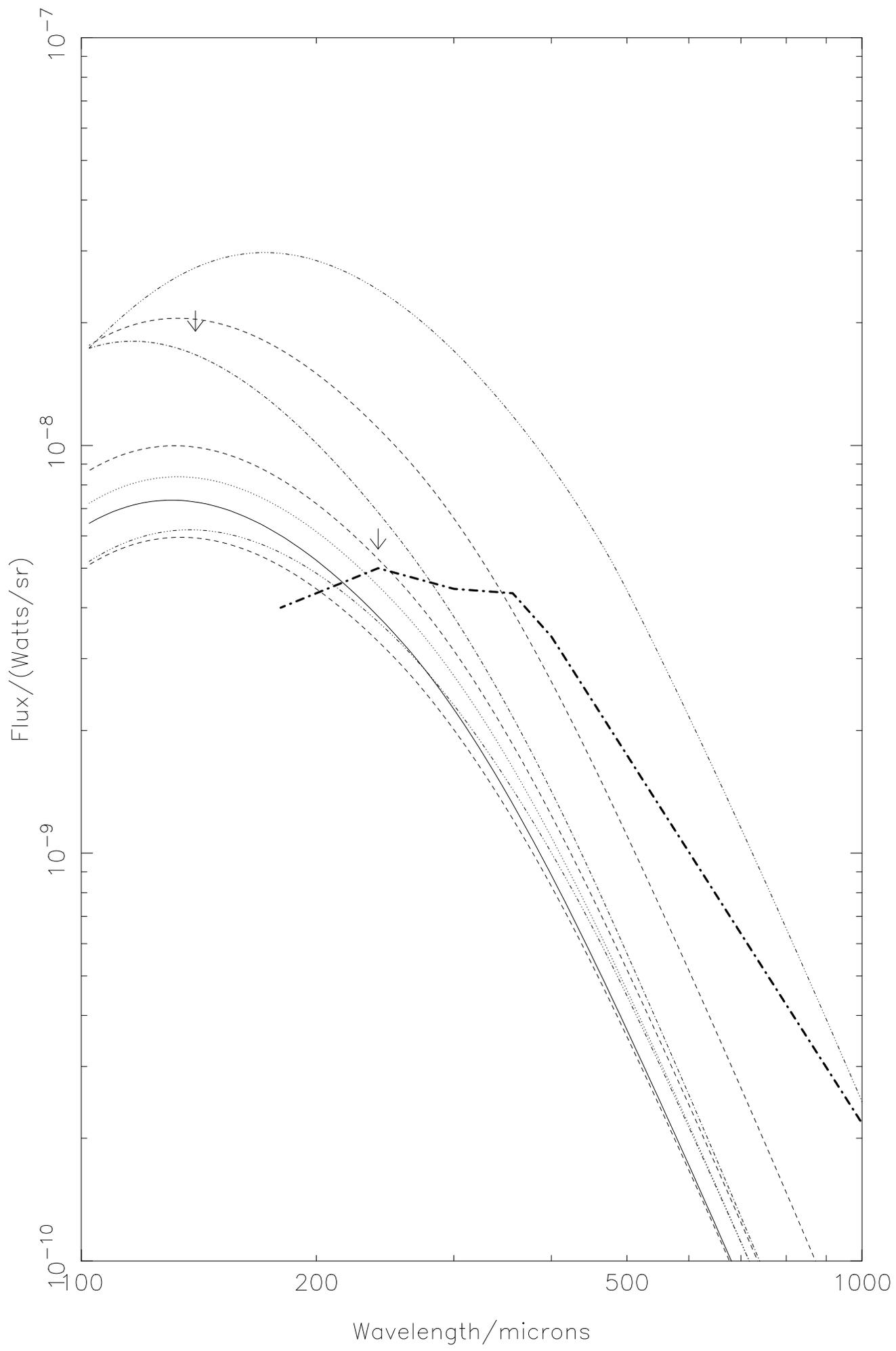
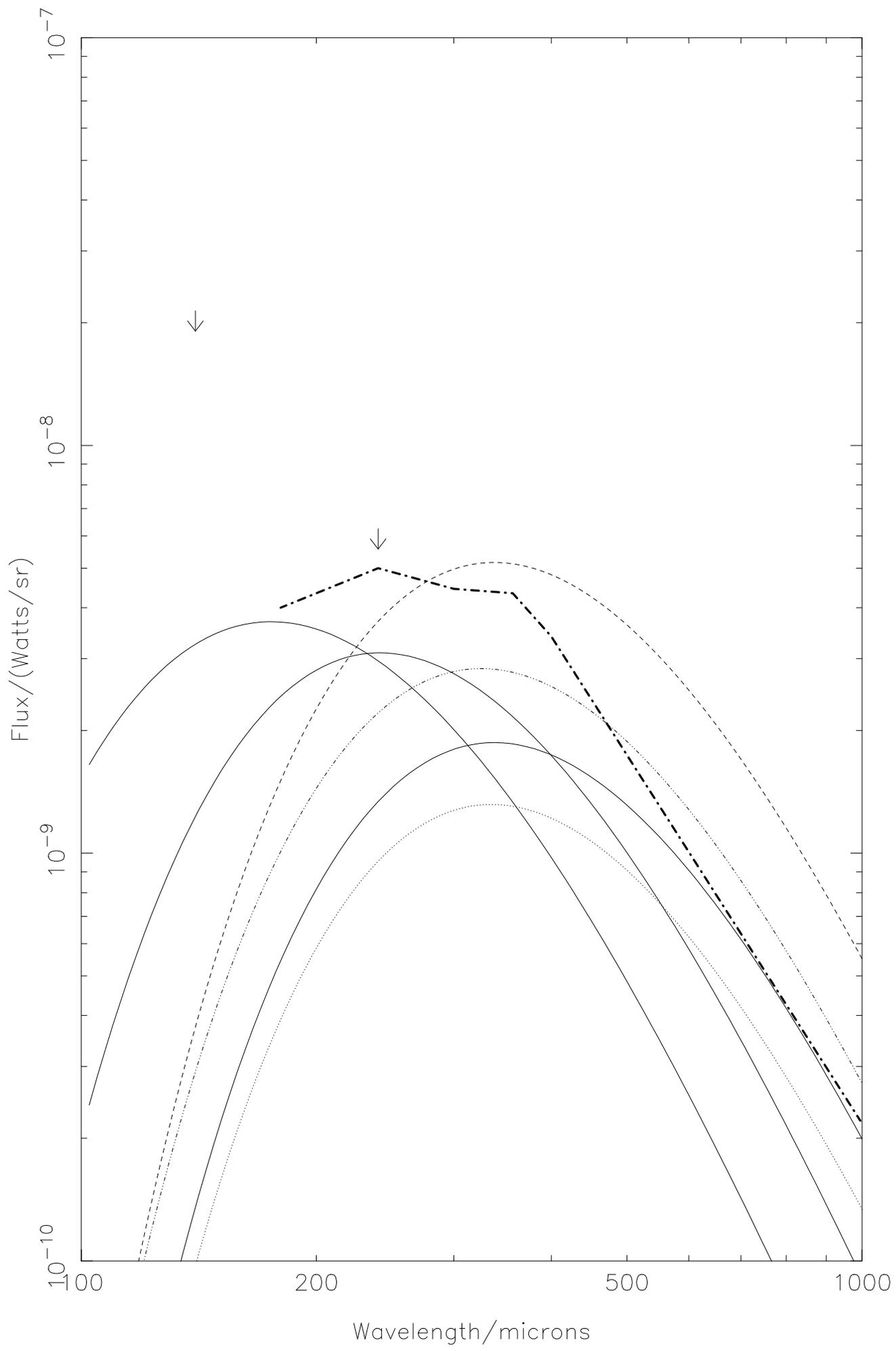
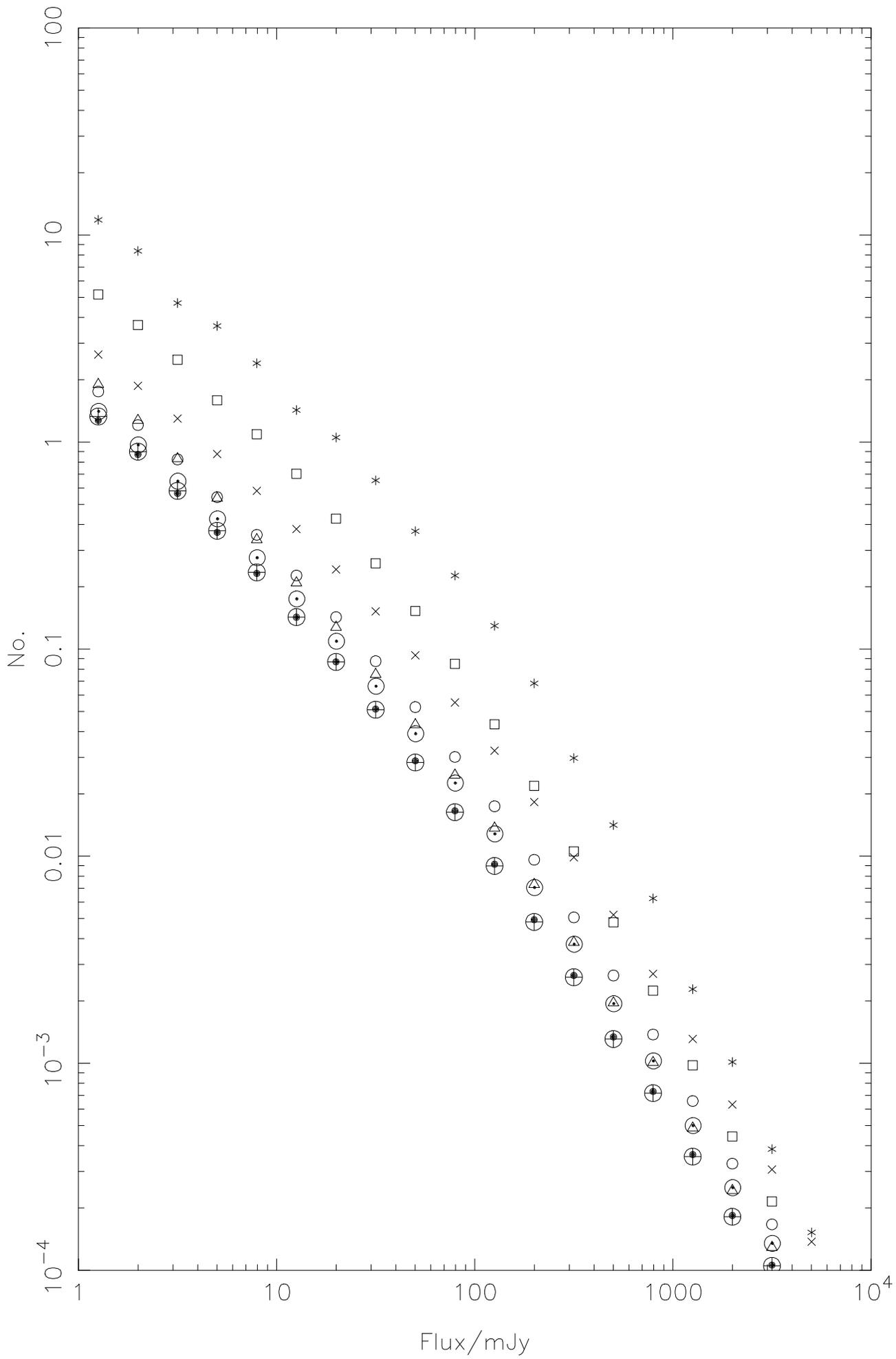
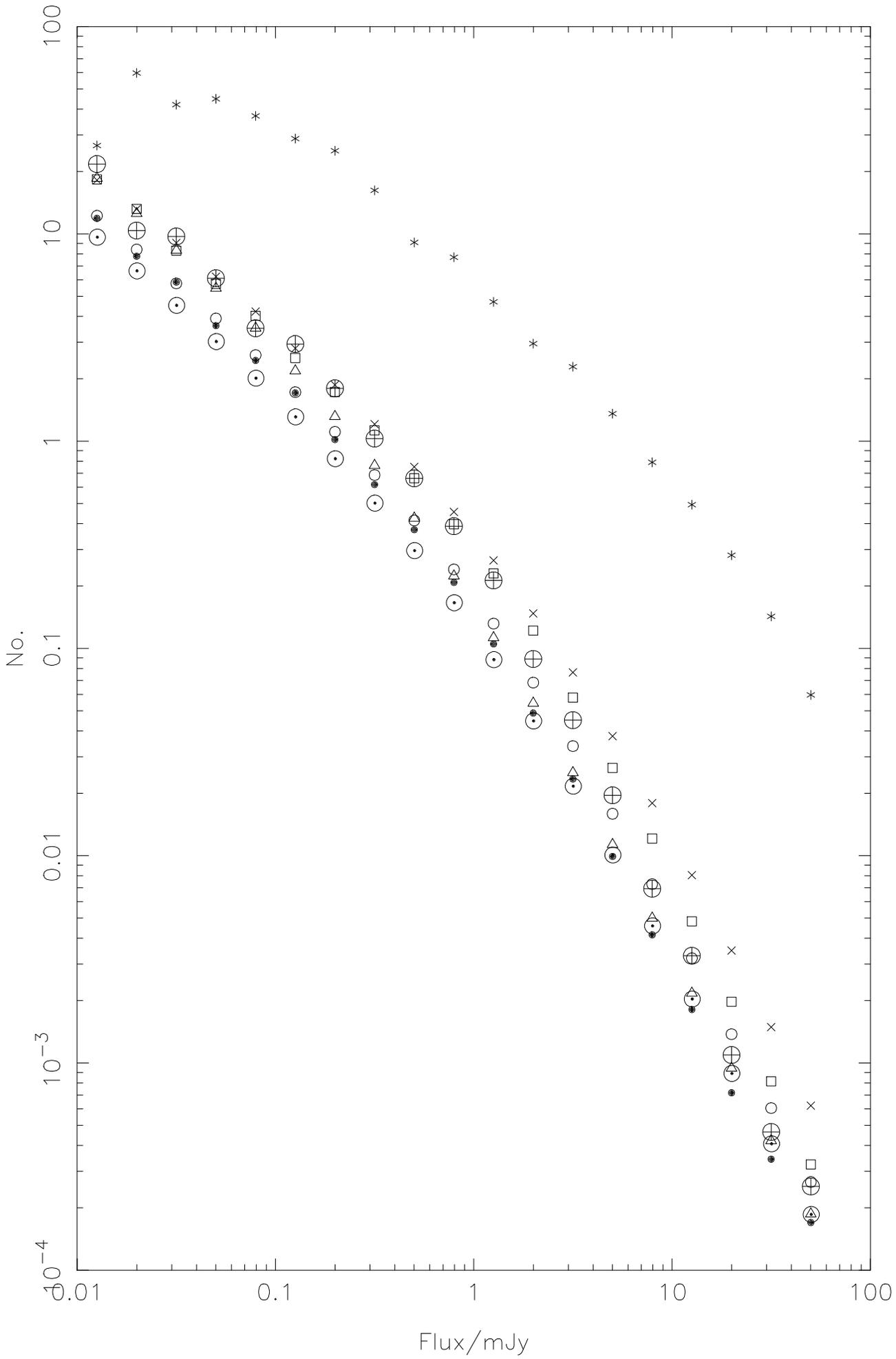
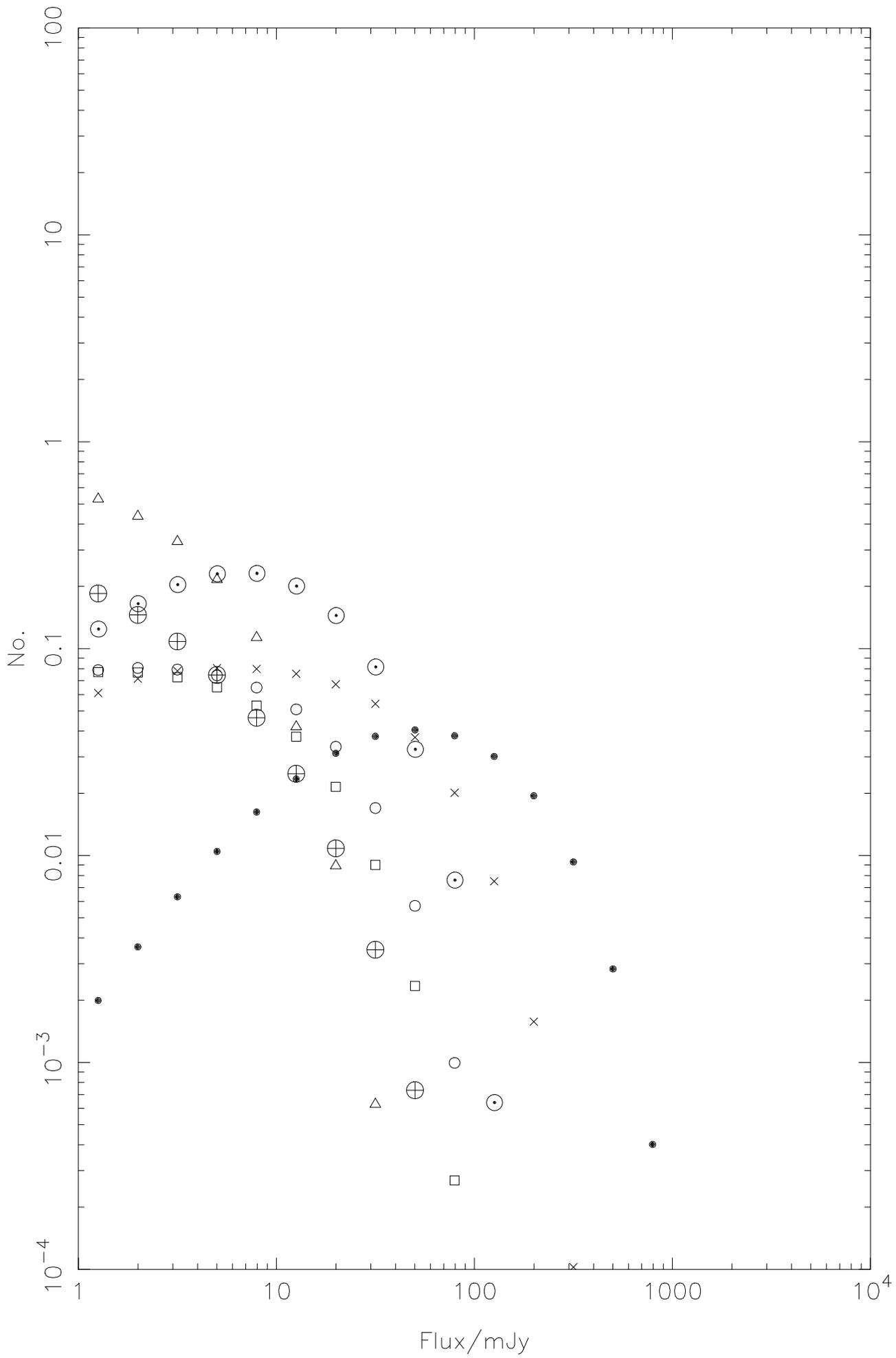
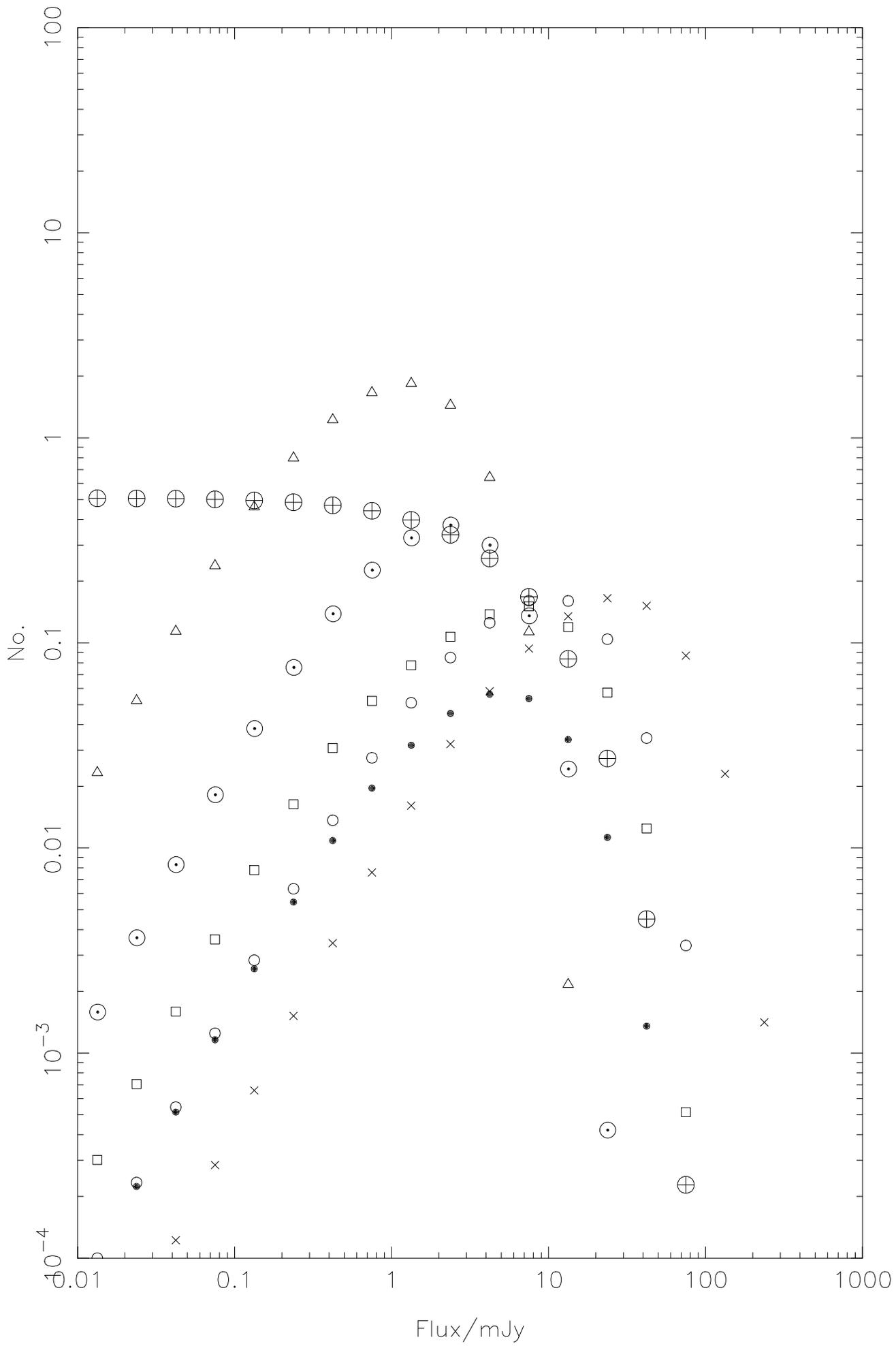
\end